# New insights of quantum Zeno phenomenon: an indirect correlation effect among subsystems

Tung-Ho Shieh[1], Kun-Yuan Wu[2], Hsiu-Fen Kao[3] and Kuan-Ming Hung[4,*]

[1]Department of Electro-Optical Engineering, Kun Shan University, Tainan City 710, Taiwan, R. O. C.

[2]Device Division, Advanced Technology and Development, United Microelectronics Co., Tainan City 710, Taiwan, R. O. C.

[3]Institute of Photonics and Communications, National Kaohsiung University of Applied Sciences, Kaohsiung City 807, Taiwan, R. O. C.

[4]Department of Electronics Engineering, National Kaohsiung University of Applied Sciences, Kaohsiung City 807, Taiwan, R. O. C.

[*]Correspondence and requests for materials should be addressed to K.M.H. (E-mail: kmhung@kuas.edu.tw).

**Frequent Measurements on an unstable particle located at observable initial state freeze the particle on this state, known as quantum Zeno effect[1-14]. Measurements on an observable subspace further open the prelude of quantum Zeno dynamics[15-17]. These phenomena affect the results of quantum measurement that has been widely used in quantum information and quantum computation[18-21]. However, this common argument is insufficient when the initial state is coupled with a continuum reservoir. In such an irreversible system, intrinsic decay property destroys the frozen behavior. Although it has been proven that the decay rate of the initial state will be affected by measurements[7], the knowledge of detailed mechanisms for this measurement-dependent decay is still limited. In this work, we found based on three-subsystem perturbation theory that the indirect correlation between a continuum reservoir and a measurement apparatus through the interaction with a quantum system causes a Rabi-like intermodulation among these subsystems. The indirect correlation effect from this combination drives the system into quantum Zeno and**

**reactivation regions, which are dependent upon the Rabi strength, pulse duty, and its repetition rate. We also found that the dynamical changes of decay rate are quite different in continuous and pulsed measurements, and, thus, significantly affect the dynamics of the system.**

Let's start from a generalized argument of Quantum Zeno Effect (QZE)[7,22]. If a particle in its initial state $|\Psi_0\rangle$ at $t=0$ is ruled by Hamiltonian $H$, the probability to find the particle in $|\Psi_0\rangle$ at $t>0$ is $p(t) = |\langle\Psi_0|e^{-iHt}|\Psi_0\rangle|^2$ (in units of $\hbar=1$), and is denoted by $p(t) = |\langle e^{-iHt}\rangle|^2$. For a finite Hamiltonian variance $\overline{V^2} = \overline{V_h^2} + \overline{V_a^2}$ ( $\overline{V_h^2} = \langle H_h^2\rangle - \langle H_h\rangle^2$ and $\overline{V_a^2} = \langle H_a^2\rangle + \langle H_a\rangle^2$ ) and anti-Hermitian potential $\overline{V_a}$ ($\overline{V_a} = 2\langle H_a\rangle$) at a short time $t$ results in $\overline{V_a}t \ll 1$ and $\overline{V^2}t^2 \ll 1$. Then the probability approaches $p(t) \cong 1 - i\overline{V_a}t - \overline{V^2}t^2$ where $H_h$ ($H_a$) is the Hermitian (anti-Hermitian) part of $H$. When N ideal measurements with time intervals $\tau$ ($t = N\tau$) are performed on the system, $p(t)$ follows

$$p(t) \cong 1 - i\overline{V_a}t - \overline{V^2}\tau t. \tag{1}$$

If the system is Hermitian ($H^+ = H$), the scattered particle has equal probability to jump forward and backward between its initial state $|\Psi_0\rangle$ and its final state $|\Psi_k\rangle$. The Hermitian characteristic results in $\overline{V_a} = 0$, $\overline{V_a^2} = 0$, and the scattered particle not changing its state when $N \to \infty$. This is known as the quantum Zeno paradox. If the quantum system is coupled to a continuum environment, the transition back to $|\Psi_0\rangle$ can be neglected, which makes the system irreversible, i.e., $H^+ \neq H$.

For a decaying state of the decay rate $\gamma$, $\overline{V_a}$ and $\overline{V_a^2}$ have the forms of $\overline{V_a} = -i\gamma$ and $\overline{V_a^2} = -\gamma^2/2$, and the probability becomes $p(t) \cong 1 - \gamma t - \overline{V_h^2} \tau t + (\gamma^2 \tau/2) t$. The particle always decays with the rate $\gamma$ from its initial state to the ground state when $N \to \infty$. This is conflicting to the argument of QZE, because regardless of whether $\gamma$ is time-dependent or not, $\gamma$ is independent of measurement[23]. Although it has been proven that the decay process can be modified by frequent observations[7], but the existence of QZE in such systems still needs to rely on a more precise mechanism to determine the relationship between the measurement and the decay rate. In this paper, indirect correlation effect (ICE) is proposed to be a specific mechanism that dominates the decay rate of quantum state.

To describe the ICE mechanism, let's consider a quantum system (QS) of self Hamiltonian $H_S$ adiabatically coupled to decoherence reservoir (DR) of $H_R$ and a measurement apparatus (MA) of $H_A$ through the interactions of $V_R$ and $V_A$, respectively (see Figure 1a). The total Hamiltonian of the system can be written as

$$H = H_0 + H_R + V_I, \qquad (2)$$

$$H_0 = H_S + H_A$$

$$V_I = e^{-\varepsilon|t|}(V_R + V_A).$$

$\varepsilon$ is an infinitesimal constant. The quantum decoherence can come from the radiation-induced deformation in the cavity photon system[24]; the phonon scattering in the solid-state system[25]; or the optical radiation in the atomic quantum system[26]. Therefore, the interacting Hamiltonian $V_R$ has the form[25]

$$V_R = \sum_{i \ne j,q} \gamma_{ijq} (A_q + A_q^+) |i\rangle\langle j|, \qquad (3)$$

where $A_q$ ($A_q^+$) is the annihilation (creation) operator of the DR's boson of mode q and the coupling strength $\gamma_{ijq}$, and $|i\rangle$ is the $i^{th}$ QS state. As well as, time-dependent interaction potential V$_A$ of QS-to-MA can be written as[25]

$$V_A = \sum_{i \neq j, k} g_{ijk}(t)\left(B_k + B_k^+\right)|i\rangle\langle j|. \tag{4}$$

$g_{ijk}$ is the coupling strength (a time-dependent envelope function) of the k mode photon coupled to the dipole moment between QS states |i> and |j>. $B_k$ ($B_k^+$) is the annihilation (creation) operator of the MA's photon of mode k. The solution space of the system is a direct product of the subspaces of QS, DR, and MA, i.e., $|\Psi(t)\rangle = |\Phi_0(t)\rangle \otimes |\Phi_R(t)\rangle$ with $|\Phi_0(t)\rangle = |\Phi_S(t)\rangle \otimes |\Phi_A(t)\rangle$.

In the interaction picture, the dynamics of the system follows the interaction Schrodinger equation

$$i\partial_t |\Psi(t)\rangle_I = \hat{V}_I(t)|\Psi(t)\rangle_I, \tag{5}$$

with $\hat{V}_I(t) = e^{i(H_0 + H_R)t} V_I e^{-i(H_0 + H_R)t}$, $|\Psi(t)\rangle_I = e^{i(H_0 + H_R)t}|\Psi(t)\rangle$. In the adiabatic approximation, the wave function at $t \to -\infty$ is static, separable, and assumed to be well known and normalized. The solution of equation (5) is

$$|\Psi(t)\rangle_I = \left(1 + \sum_{n=1}^{\infty}(-i)^n \prod_{m=1}^{n}\int_{-\infty}^{t_{m-1}} dt_m \hat{V}_I(t_m)\right)|\Psi(-\infty)\rangle_I, \tag{6}$$

with $t_0 = t$. By taking an average $_I\langle\Phi_R(-\infty)|$ on equation (6) with respect to DR, one can obtain the dressed wave function $|\Phi_0(t)\rangle_I \equiv {}_I\langle\Phi_R(-\infty)|\Psi(-\infty)\rangle_I$ of H$_0$

$$|\Phi_0(t)\rangle_I = \left(1 + \sum_{n=1}^{\infty}(-i)^n {}_I\!\left\langle\Phi_R(-\infty)\left|\prod_{m=1}^{n}\int_{-\infty}^{t_{m-1}} dt_m \hat{V}_I(t_m)\right|\Phi_R(-\infty)\right\rangle_I\right)|\Phi_0(-\infty)\rangle_I. \tag{7}$$

For non-degenerate states of DR, the average $_I\langle\Phi_R(-\infty)|\hat{O}(t)|\Phi_R(-\infty)\rangle_I$ of operator $\hat{O}(t)$ is exactly its ensemble average denoted by $\langle\hat{O}(t)\rangle_R$.

The derivation is based on the following reasons. First, the statistic ensemble of stochastic operators of DR at thermal equilibrium will be zero, i.e., $<A_q>_R = <A_q^+>_R = 0$. Second, $g \gg \gamma$. Third, Wick's theorem is used to decompose the ensemble averages of four or more DR-boson operators into the product of pairing averages[25]. Finally, since the ensemble average only removes the DR's degrees of freedom but leaves the unsolved operators of $H_0$, the time order of the QS-to-DR and QS-to-MA interactions is important and severely affects the decaying behaviors. When quantum measurements are performed on QS, the measurement not only provides the information of QS but also disturbs its environment. This disturbance will be fed back from the environment at a later time when it will affect the QS itself through the processes as shown in Figure 1**b**. Of course, all decoherence processes of any measured photon number are possible, and should estimate their impacts on decay and decoherence ─ a Rabi-like modulation. It is much different from the existing theories that only individually deal the impacts of DR and MA on QS[23,27]. Consequently, we collect all the sequences of the irreducible diagrams in the perturbation expansion, which implies the following equivalent Schrodinger equation (detailed derivations see Suppl. A)

$$i\partial_t |\Phi_0(t)\rangle = H_{red}(t)|\Phi_0(t)\rangle. \tag{8}$$

$H_{red} = H_0 + V_{red}(t)$ is the reduced Hamiltonian. $V_{red}(t) = V_d(t) + V_{od}(t)$ is the reduced interaction potential in self-consistent form with the diagonal part

$$V_d(t) = -ie^{-iH_0 t}\left\langle \hat{V}_R(t)\int_{-\infty}^{t} dt_1 \hat{V}_R(t_1)\right\rangle_R e^{iH_0 t_1}\hat{U}(t_1,t) + X_0(t), \tag{9}$$

and the off-diagonal part

$$V_{od}(t) = V_A + X_1(t), \tag{10}$$

where

$$X_k(t) = \sum_{n=1}^{\infty} (-i)^{2n+1+k} e^{-iH_0 t} \left\langle \hat{V}_R(t) \left( \prod_{m=1}^{2n+k} \int_{-\infty}^{t_{m-1}} dt_m \hat{V}_A(t_m) \right) \int_{-\infty}^{t_{2n+k}} dt_{2n+1+k} \hat{V}_R(t_{2n+1+k}) \right\rangle_R,$$

$$\times e^{iH_0 t_{2n+1+k}} \hat{U}(t_{2n+1+k}, t)$$

$k = 0$ or $1$, and $t_0 = t$. $\hat{U}(t,t')$ is the time-development operator defined in Suppl. (A-14). The potential $V_d$ gives rise to energy decay while $V_{od}$ dominates Rabi transition and phase decoherence. Notably, $X_k(t)$ in the equations (9) and (10) caused by ICE provides an intermodulation channel among DR and MA.

Now we can apply the above theory to a two-level system with the interacting potential $V_A = -g(t)e^{i\omega t}|0\rangle\langle 1| - g(t)e^{-i\omega t}|1\rangle\langle 0|$ (within rotating-wave approximation), where $\omega$ is the photon frequency and $2g$ is the Rabi frequency. In order to analytically solve equations (9) and (10), we only consider the lowest-order diagonal correction in the exponent of $\hat{U}$, i.e. $\hat{U}(t,t') \cong e^{\left(-iH_0 - \sum_i \gamma_i |i\rangle\langle i|\right)(t-t')}$, because the lowest-order off-diagonal terms are negligible for $\omega \gg g$. $\gamma_i$ is the decay rate of the i$^{th}$ energy level of QS, and substitutes the role of ε in adiabatically booting process. For N+1 square-pulse measurements, the coupling strength $g(t)$ can be modeled as $g(t) = \sum_{n=0}^{N} \lambda[\theta(t-t_n) - \theta(t-t_n-\tau_{on})]$, with the pulse height $\lambda = \Lambda/2$ ($\Lambda$ is the Rabi frequency), the pulse width $\tau_{on}$, and the pulse starting time $t_n$ of the n$^{th}$ pulse. $\theta(t)$ is the unit step function. Equations (9) and (10) have analytical solutions for the cases of resonance $\omega = \Omega$ and large detuning $|\omega - \Omega| \gg \gamma_i$, where $\Omega$ is the energy level of $|1\rangle$ being measured with respect to $|0\rangle$. In resonance, equations (9) and (10) are calculated to be (seen Suppl. B)

$$V_d = -i \sum_{j=0}^{1} \gamma_j \left\{ 1 + \sum_{n=0}^{N} [K_{d,j}(\tau_n, \tau_n)\Theta_{n,1}(t) + K_{d,j}(\tau_{on}, \tau_n)\Theta_{n,0}(t)] \right\} |j\rangle\langle j| \quad (11)$$

and

$$V_{od} = V_A + \sum_{j=0}^{1} \gamma_j e^{(-1)^{1-j} i\Omega t} \sum_{n=0}^{N} \left[ K_{o,j}(\tau_n, \tau_n) \Theta_{n,1}(t) + K_{o,j}(\tau_{on}, \tau_n) \Theta_{n,0}(t) \right] |1-j\rangle\langle j|, \quad (12)$$

where

$$K_{d,j}(x,y) = \frac{\lambda e^{-\gamma_j y}}{\gamma_j^2 + \lambda^2} \left[ \lambda \cos(\lambda x) + \gamma_j \sin(\lambda x) - \lambda e^{\gamma_j x} \right] \theta(y),$$

$$K_{o,j}(x,y) = \frac{\lambda e^{-\gamma_j y}}{\gamma_j^2 + \lambda^2} \left[ \gamma_j e^{\gamma_j x} + \lambda \sin(\lambda x) - \gamma_j \cos(\lambda x) \right] \theta(y),$$

$\gamma_1 = \frac{\gamma}{2}[1 + N_B(\Omega)]$ , $\gamma_0 = \frac{\gamma}{2} N_B(\Omega)$ ($N_B$ is the Bose-Einstein distribution),

$\gamma = 2 \text{Im} \left( \sum_q \gamma_q^2 \frac{1}{\Omega - \omega_q - i\varepsilon} \right)$, $\tau_n = t - t_n$, and $\Theta_{n,m}(t) = m\theta(\tau_n) + (-1)^m \theta(\tau_n - \tau_{on})$.

$\Theta_{n,1}(t_n \le t \le t_n + \tau_{on})$ and $\Theta_{n,0}(t > t_n + \tau_{on})$ equal one, otherwise both of them equal zero. Since the intermodulation effect in the off-diagonal part (i.e., $V_{od}$) modifies the counter-rotating strength, the correction in equation (12) is negligible in the rotating-wave approximation, i.e. $V_{od} \cong V_A$.

In large detuning, we obtain

$$V_d = -i \sum_{j=0}^{1} \gamma_j \left\{ 1 + \sum_{n=0}^{N} \left[ \Xi_j(\tau_n, \tau_n) \Theta_{n,1}(t) + \Xi_j(\tau_{on}, \tau_n) \Theta_{n,0}(t) \right] \right\} |j\rangle\langle j|, \quad (13)$$

$$\Xi_j(x,y) = \frac{\lambda^2 e^{-\gamma_j y}}{\lambda^2 + (-1)^j i \gamma_j \delta\omega} \left( e^{(-1)^j i\lambda^2 x/\delta\omega} - e^{\gamma_j x} \right) \theta(y)$$

with $\delta\omega = \omega - \Omega$. For the same reason as that in resonance case, the off-diagonal potential of $V_{od}$ is approximated to $V_A$.

In numerical estimation, the average potential $\overline{V_d^j(t)}$ of the quantum state $|j\rangle$ can be written as $\overline{V_d^j(t)} = \Sigma_j(t) - i\frac{1}{2}\Re_j(t)$, where $\Sigma_j$ and $\frac{-1}{2}\Re_j$ are the real and the imaginary parts of $\langle j|V_d(t)|j\rangle$, respectively. $\Re_j$ is the decay rate; $\Sigma_j$ is the ICE-induced self energy that renormalizes the bare energy of quantum state $|j\rangle$ and is zero in resonance case. For simplicity, the temperature of the system is set to zero, and thus $\overline{V_d^0} = 0$. The calculated results are respectively plotted in Figures 2 and 3 for the cases of resonance and large detuning.

There are three noticeable features in these figures. First, continuous measurement (CM) and pulsed measurement (PM) as plotted in Figures 2a and 3a show different behaviors. In CM, for $t \to \infty$, the quantum system with ICE goes into a non-decay region, i.e., $\Re_1(t \to \infty) \to 0$, because of the memory effect, rather than retaining a constant which is similar to the case without ICE. Second, these figures exhibit the ICE-induced intermodulation effect on the decay rate (see solid lines in Figures 2**a** and 3**a**) and self energy (see dashed line in Figure 3**a** and see Figure 3**b**) of DR. Modulated periods for resonance and large detuning are $\lambda$ and $\lambda^2/|\delta\omega|$ as respectively given by equations (11) and (13). Third, in PM, both $\Sigma_j$ and $\Re_j$ are driven away from their initial value towards negative one by MA-field pulse during pulse-on $\tau_{on}$, and exponentially recovered back to their initial value during pulse-off $\tau_{off}$. If $\tau_{off}$ is sufficiently short such that $\Sigma_j$ and $\Re_j$ cannot be completely recovered before the next incoming field pulse, a reverse accumulation effect occurs (see Figures 2**a** and 3**a**). The accumulated value is a function of $\tau_{on}$, $\lambda$, and the duty cycle D ($D = \tau_{on}/\tau$, $\tau = \tau_{on} + \tau_{off}$) of a sequence field pulses (see Figures 2**b**, 2**c**, and 3**c**). As the number N of field pulses increases, the decay rate $\Re_1$ decreases towards a saturated value. When $\lambda'\tau_{on} \ll 1$, the saturated value that can be extracted from equations (11) and (13) has the form $\Re_1 \cong 2\gamma_1 - \lambda'^2 \tau_{on} D$ with $\lambda' = \lambda$ (for $\delta\omega = 0$) and $\lambda' = \lambda^2/|\delta\omega|$ (for $|\delta\omega| \gg \gamma$). Since the QZE occurs when $\Re_1 = 0$, we obtain the QZE-appearance criterion of a pulsed-on time $\tau_{on} \cong \frac{2\gamma_1}{\lambda'^2 D}$ (black lines in Figures 2**b**, 2**c**, 3**b**, and 3**c**). If $\tau_{on} > \frac{2\gamma_1}{\lambda'^2 D}$ (i.e., $\Re_1 < 0$), the decaying quantum state $|1\rangle$ is driven into reactivation region. For a small $\tau_{on}$ and/or a small D, the system is slightly affected by PM because both $\Sigma_j$ and $\Re_j$ of the system have a small change in these situations.

Based on our theory and simulation results, we conclude that the ICE between DR and MA through the coupling to QS significantly modulate their coupling strengths. Especially, in PM, this effect could drive the system into quantum Zeno and reactivation regions. We believe that the ICE needs to be taken into account when considering any multiple quantum system because ICE significantly impacts its quantum dynamics. In addition, the quantum system exhibits different dynamics in PM and CM. In PM, the decay process is strongly dependent on the measured conditions such as $\lambda$, $\delta\omega$, $D$, and $\tau_{on}$ (or its repetition rate), while in CM this process is solely controlled by $\lambda$. The phenomena described herein are fundamental in the quantum mechanics, and have to be considered in relevant applications.

**Acknowledgements**

This work is supported by the Ministry of Science and Technology, R. O. C., under the contract NSC100-2112-M-151-001-MY3.

**Author contributions**

K.M.H. proposed the ideas. All authors have equal contributions to the other work.

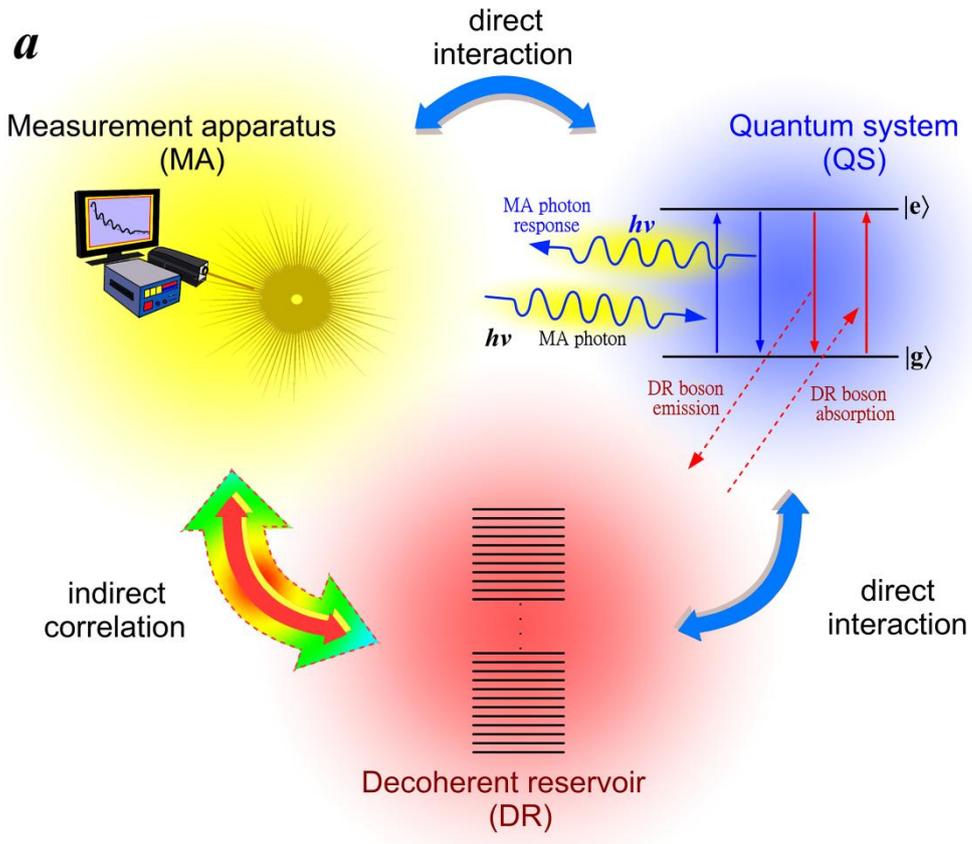

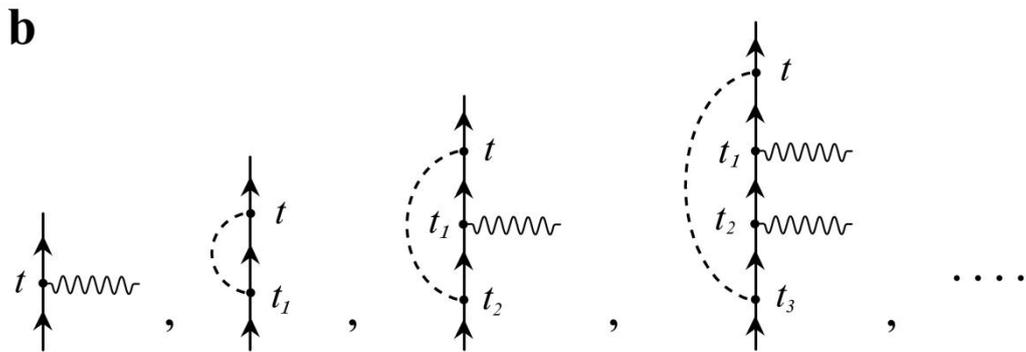

**Figure 1 a**, Schematic plot of the interaction between three subsystems. **b**, Plot of the irreducible diagrams considered in this work.

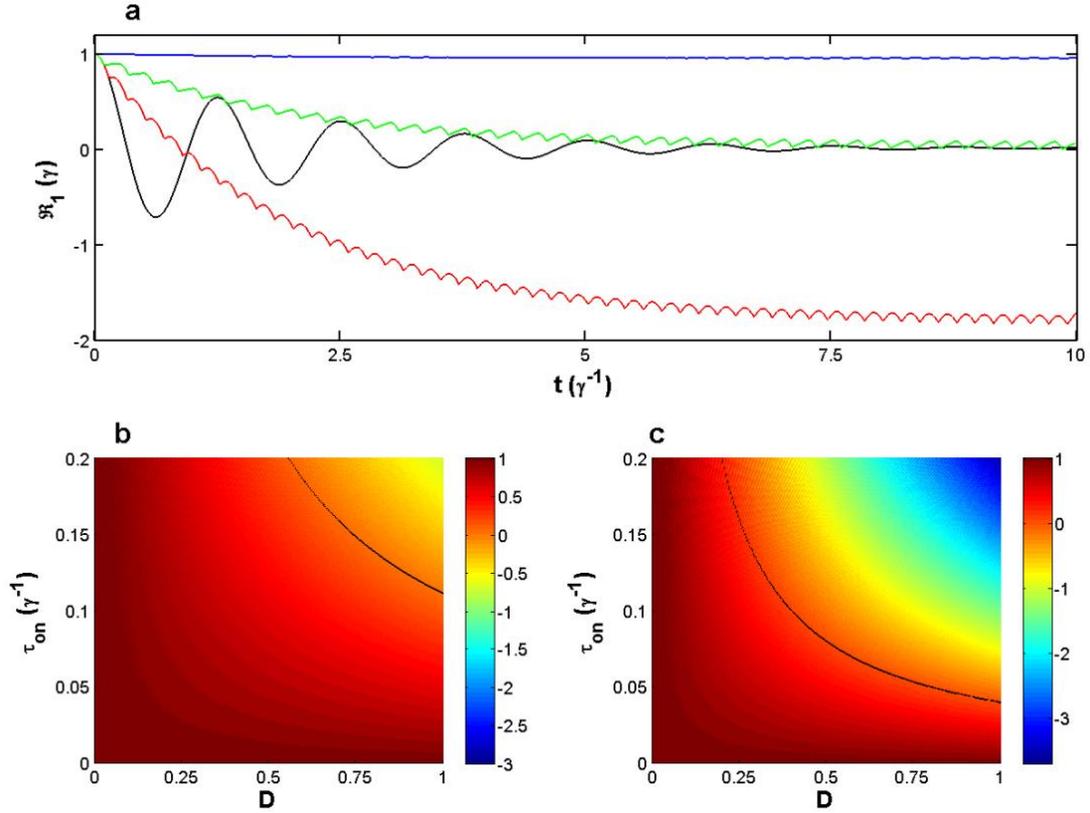

**Figure 2** The decay rate $\mathfrak{R}_1(t)$ for $\delta\omega = 0$. The self energy $\Sigma_j$ disappears in resonance. **a**, Plot $\mathfrak{R}_1(t)$ for $\lambda = 5\gamma$ (CM: black line; PM for $D = 0.1$ and $\tau_{on} = 0.02\gamma^{-1}$: blue line; PM for $D = 0.4$ and $\tau_{on} = 0.1\gamma^{-1}$: green line; PM for $D = 0.8$ and $\tau_{on} = 0.15\gamma^{-1}$: red line). **b**, Plot $\mathfrak{R}_1(t = 20\gamma^{-1})$ as a function of $D$ and $\tau_{on}$ for $\lambda = 3\gamma$. **c**, Plot $\mathfrak{R}_1(t = 20\gamma^{-1})$ as a function of $D$ and $\tau_{on}$ for $\lambda = 5\gamma$. The black lines in **b** and **c** show the QZE condition, and the area above these lines is reactivation region.

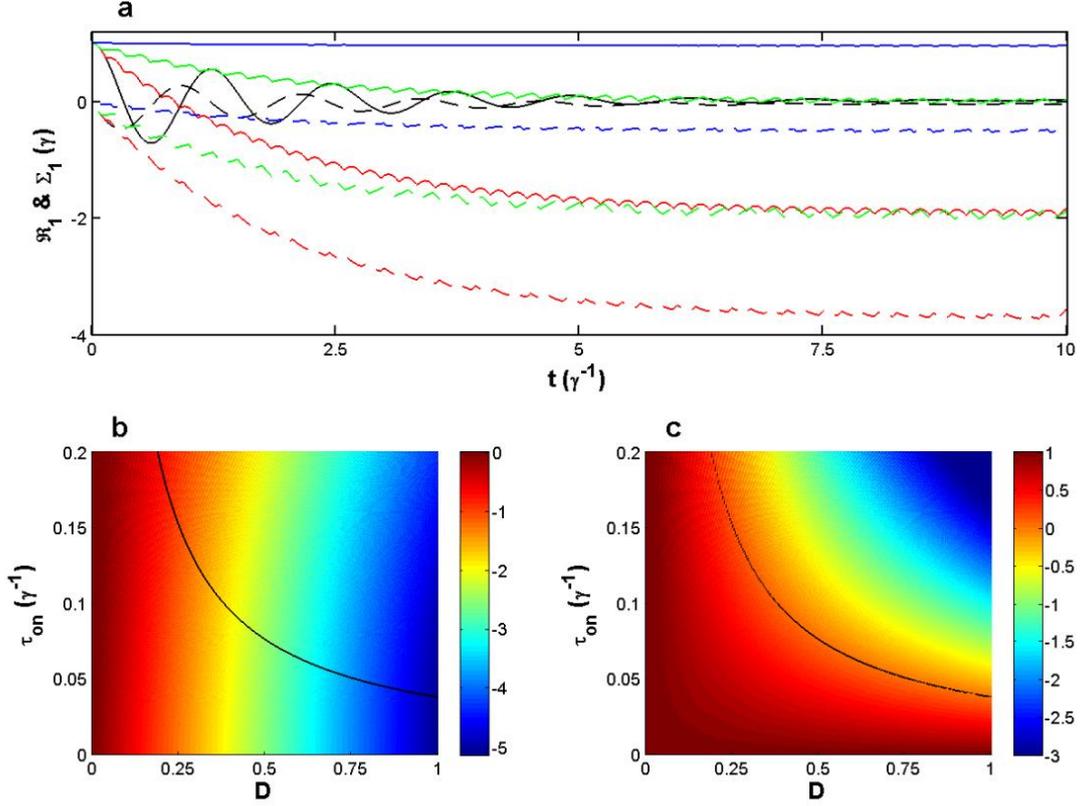

**Figure 3** The decay rate $\Re_1(t)$ and the self energy $\Sigma_1(t)$ for $|\delta\omega| \gg \lambda$. **a**, Plot $\Re_1(t)$ (solid lines) and $\Sigma_1(t)$ (dashed lines) for $\lambda = 16\gamma$ and $\delta\omega = 50\gamma$ (CM: black line; PM for $D = 0.1$ and $\tau_{on} = 0.02\gamma^{-1}$: blue line; PM for $D = 0.4$ and $\tau_{on} = 0.1\gamma^{-1}$: green line; PM for $D = 0.8$ and $\tau_{on} = 0.15\gamma^{-1}$: red line). In PM the energy of quantum state $|1\rangle$ is renormalized to increase the detuning energy $|\delta\omega|$, while in CM the energy exhibits a sinusoidal oscillation with exponentially decaying to zero. The phase of $\Sigma_1$ is 90 degrees ahead of $\Re_1$. **b,** Plot $\Sigma_1(t = 20\gamma^{-1})$ as a function of $D$ and $\tau_{on}$ for $\lambda = 16\gamma$ and $\delta\omega = 50\gamma$. **c,** Plot $\Re_1(t = 20\gamma^{-1})$ as a function of $D$ and $\tau_{on}$ for $\lambda = 16\gamma$ and $\delta\omega = 50\gamma$. The black lines in **b** and **c** show the QZE condition, and the area above these lines is reactivation region.

## Supplementary Information

**Suppl. A: Derivation of three-subsystem perturbation theory**

In theory, to consider ICE gives rise to two key issues. First, the DR usually has infinite degrees of freedom, which complicates the system, especially, in the case of multiple quantum system. Second, in quantum information application, a quantum system requires to maintain a longer coherent time in comparison to manipulated or measured time, i.e., the interacting strength of QS-to-MA is much stronger than QS-to-DR. Therefore, in the theoretical treatment of ICE, the former must be treated as a strong coupling. This increases the difficulty of dealing with perturbation theory. Here, we derive a generic three-subsystem perturbation theory with consideration of ICE.

Since the statistics ensemble of stochastic operators of DR at thermal equilibrium is zero, i.e. $<A_k>_R = <A_k^+>_R = 0$, the expansion of equation (7) just includes even order terms with DR-boson operators. Then, we apply Wick's theorem to decompose the ensemble averages of four or more DR-boson operators into the product of pairing averages and retain the lowest order terms according to $g \gg \gamma$, i.e., for example,

$$\overline{\left\langle \hat{A}_{q1}\hat{A}_{q2}^+\hat{A}_{q3}\hat{A}_{q4}^+ \right\rangle}_R$$
$$= \delta_{q1,q2}\delta_{q3,q4}\overline{\left\langle \hat{A}_{q1}\hat{A}_{q2}^+ \right\rangle}_R \overline{\left\langle A_{q3}A_{q4}^+ \right\rangle}_R + \delta_{q1,q4}\delta_{q2,q3}\overline{\left\langle \hat{A}_{q1}\hat{A}_{q2}^+\hat{A}_{q3}\hat{A}_{q4}^+ \right\rangle}_R . \quad \text{(A-1)}$$
$$\cong \delta_{q1,q2}\delta_{q3,q4}\overline{\left\langle \hat{A}_{q1}\hat{A}_{q2}^+ \right\rangle}_R \overline{\left\langle A_{q3}A_{q4}^+ \right\rangle}_R$$

The first term in equation (A-1) does not belong to irreducible group because it is expandable by the second irreducible diagram in Figure 1**b**. Since the integrations in time order have to be retained, the second term in equation (A-1) cannot be decomposed and belongs to fourth-order irreducible diagrams. Thus, this term can be neglected for $g \gg \gamma$. This implies the following equation

$$\overline{\left\langle V_{R1}V_{R2}V_{R3}V_{R4} \right\rangle}_R = \delta_{R1,R2}\delta_{R3,R4}\overline{\left\langle \hat{V}_{R1}\hat{V}_{R2} \right\rangle}_R \overline{\left\langle \hat{V}_{R3}\hat{V}_{R4} \right\rangle}_R$$
$$+ \delta_{R1,R4}\delta_{R2,R3}\overline{\left\langle \hat{V}_{R1}\hat{V}_{R2}\hat{V}_{R3}\hat{V}_{R4} \right\rangle}_R + \delta_{R1,R3}\delta_{R2,R4}\overline{\left\langle \hat{V}_{R1}\hat{V}_{R2}\hat{V}_{R3}\hat{V}_{R4} \right\rangle}_R \quad . \quad \text{(A-2)}$$

In Feynman diagram representation, equation (A-2) corresponds to the following diagram

$$\overline{\langle V_{p1}V_{p2}V_{p3}V_{p4}\rangle}_R = \quad\text{(diagram)} + \text{(diagram)}$$

$$+ \text{(diagram)},$$

where the DR-boson line (dashed line) includes the emission and absorption processes, i.e.,

$$\text{(diagram)} = \text{(diagram)} + \text{(diagram)}.$$

For the same reason as discussed in equation (A-1), the second and third terms in equation (A-2) are ignorable for $g \gg \gamma$. With collecting all orders of diagrams that are expandable by the irreducible group in Figure 1**b**, equation (7) can be written as

$$|\Phi_0(t)\rangle_I = \left\{1 + \int_{-\infty}^t dt_1 (-i)\hat{f} \bullet \left[1 + (-i)\hat{f} + (-i)^2 \hat{f} \bullet \hat{f} + \cdots\right]\right\}|\Phi_0(-\infty)\rangle. \tag{A-3}$$

$\hat{f}$, included all irreducible diagrams, is the super-operator defined as

$$\begin{aligned}\hat{f} &= \overline{\langle \hat{V}_{A1}\rangle}_R + (-i)\overline{\langle \hat{V}_{R1}\hat{V}_{R2}\rangle}_R + (-i)^2 \overline{\langle \hat{V}_{R1}\hat{V}_{A2}\hat{V}_{R3}\rangle}_R \\ &+ (-i)^3 \overline{\langle \hat{V}_{R1}\hat{V}_{A2}\hat{V}_{A3}\hat{V}_{R4}\rangle}_R + (-i)^4 \overline{\langle \hat{V}_{R1}\hat{V}_{A2}\hat{V}_{A3}\hat{V}_{A4}\hat{V}_{R5}\rangle}_R + \cdots\end{aligned} \tag{A-4}$$

with $\overline{\langle \hat{O}_1 \cdots \hat{O}_n \rangle}_R \equiv \int_{-\infty}^{t_1} dt_2 \cdots \int_{-\infty}^{t_{n-1}} dt_n \langle \Phi_R(-\infty)|\hat{O}(t_1)\cdots \hat{O}(t_n)|\Phi_R(-\infty)\rangle$. We show hereafter that $\hat{f}$ is the reduced interaction Hamiltonian of dressed system $H_0$. If $\hat{A}$ and $\hat{B}$ are given by

$$\hat{A}(t_{a1}|t_{a2},\cdots,t_{an}) = \int_{-\infty}^{t_{a1}} dt_{a2} \cdots \int_{-\infty}^{t_{a(n-1)}} dt_{an} \langle \hat{V}_{a1}(t_{a1})\cdots \hat{V}_{an}(t_{an})\rangle_R,$$

$$\hat{B}(t_{b1}|t_{b2},\cdots,t_{bm}) = \int_{-\infty}^{t_{b1}} dt_{b2} \cdots \int_{-\infty}^{t_{b(m-1)}} dt_{bm} \langle \hat{V}_{b1}(t_{b1})\cdots \hat{V}_{bm}(t_{bm})\rangle_R,$$

where $t_{a2},\cdots,t_{an}$ and $t_{b2},\cdots,t_{bm}$ are dummy variables, then the operation $\bullet$ is defined as

$$\begin{aligned}&\hat{A}(t_{a1}|t_{a2},\cdots,t_{an}) \bullet \hat{B}(t_{b1}|t_{b2},\cdots,t_{bm}) \\ &\equiv \int_{-\infty}^{t_{a1}} dt_{a2} \cdots \int_{-\infty}^{t_{a(n-1)}} dt_{an} \langle \hat{V}_{a1}(t_{a1})\cdots \hat{V}_{an}(t_{an})\rangle_R \\ &\times \int_{-\infty}^{t_{an}} dt_{b1} \cdots \int_{-\infty}^{t_{b(m-1)}} dt_{bm} \langle \hat{V}_{b1}(t_{b1})\cdots \hat{V}_{bm}(t_{bm})\rangle_R\end{aligned} \tag{A-5}$$

and $\hat{A} \bullet c = c\hat{A}$ for constant c. Taking time derivative for equation (A-3), yields

$$i\partial_t |\Phi_0(t)\rangle_I = \hat{f} \bullet \left[1 + (-i)\hat{f} + (-i)^2 \hat{f} \bullet \hat{f} + \cdots\right] |\Phi_0(-\infty)\rangle_I. \quad (A\text{-}6)$$

In order to simplify equation (A-6), we define a time ordered operator as

$$\hat{T}\left(\int_{-\infty}^t \hat{A}\, dt_{a1} \int_{-\infty}^t \hat{B}\, dt_{b1}\right) \equiv \theta(t_{an} - t_{b1})\int_{-\infty}^t dt_{a1}\hat{A} \bullet \hat{B} + \theta(t_{bm} - t_{a1})\int_{-\infty}^t dt_{b1}\hat{B} \bullet \hat{A}. \quad (A\text{-}7)$$

Based on this definition, one can show that $\frac{1}{2!}\hat{T}\left[\left(\int_{-\infty}^t \hat{A}\, dt_{a1}\right)^2\right] = \int_{-\infty}^t \hat{A}(t_{a1},...,t_{an}) \bullet \hat{A}(t_{a1'},...,t_{an'})\, dt_{a1}$,

$\frac{1}{3!}\hat{T}\left[\left(\int_{-\infty}^t \hat{A}\, dt_{a1}\right)^3\right] = \int_{-\infty}^t \hat{A}(t_{a1},...,t_{an}) \bullet \hat{A}(t_{a1'},...,t_{an'}) \bullet \hat{A}(t_{a1''},...,t_{an''})\, dt_{a1}$, ... etc. According to these

relations we obtain

$$i\partial_t |\Phi_0(t)\rangle_I = \hat{f}(t) |\Phi_0(t_\bullet)\rangle_I, \quad (A\text{-}8)$$

where $|\Phi_0(t_\bullet)\rangle_I = \hat{U}_I(t_\bullet, -\infty) |\Phi_0(-\infty)\rangle_I$, $t_\bullet$ is the last one integral time variable of each term in the

operator $\hat{f}$ prior to $\hat{U}_I$, and

$$\hat{U}_I(t_1, t_2) \equiv \hat{T}\left(e^{-i\int_{t_2}^{t_1} \hat{f}(t')dt'}\right) \quad (A\text{-}9)$$

is the time development operator that follows the relations $\hat{U}_I(t_2, t_1) = \hat{U}_I^+(t_1, t_2)$ and

$\hat{U}_I(t_1, t_2) = \hat{U}_I(t_1, t')\hat{U}_I(t', t_2)$. According to these relations and $|\Phi_0(t)\rangle_I = e^{iH_0 t}|\Phi_0(t)\rangle$, equation (A-8)

can be written as the reduced interaction Schrodinger equation

$$i\partial_t |\Phi_0(t)\rangle_I = \hat{f}(t)\hat{U}_I^+(t, t_\bullet) |\Phi_0(t)\rangle_I, \quad (A\text{-}10)$$

which gives the following non-Markovian Schrodinger equation

$$i\partial_t |\Phi_0(t)\rangle = H_{red} |\Phi_0(t)\rangle, \quad (A\text{-}11)$$

with $H_{red} = H_0 + V_{red}$, $V_{red} = V_d + V_{od}$. The diagonal part $V_d$ and off-diagonal part $V_{od}$ of $V_{red}$

have the forms

$$V_d = e^{-iH_0 t} \hat{f}_{even}(t) e^{iH_0 t} \bullet \hat{U}(t_\bullet, t), \quad (A\text{-}12)$$

$$V_{od} = e^{-iH_0 t} \hat{f}_{odd}(t) e^{iH_0 t} \bullet \hat{U}(t_\bullet, t), \quad (A\text{-}13)$$

$$\hat{U}(t_\bullet,t) = e^{-iH_0 t_\bullet} \hat{U}_I^+(t,t_\bullet) e^{iH_0 t}. \tag{A-14}$$

$\hat{f}_{even}$ ($\hat{f}_{odd}$) is the even- (odd-) order part of $\hat{f}$. The explicit forms of $V_d$ and $V_{od}$ are given by equations (9) and (10) in the text, respectively.

**Suppl. B: Analytical solutions for two-level system**

Before the calculations of $V_d$ and $V_{od}$, we first estimate the exponent terms of $\hat{U}_I$. Since equations (9) and (10) have self-consistent form, $V_d$ and $V_{od}$ are difficult to find their analytical solutions. In order to obtain the analytical solutions, we simply keep the lowest-order terms in the exponent $(-i\int_t^{t_\bullet} \hat{f}(t')dt')$ of $\hat{U}_I$. In this situation, for $\omega \gg \lambda$, the first off-diagonal integration term $-i\int_t^{t_\bullet} \hat{V}_A(t')dt'$ is negligible, and, thus, the exponent of $\hat{U}_I$ can be approximately written as $-i\int_t^{t_\bullet} \hat{f}(t')dt' \cong (-i)^2 \int_t^{t_\bullet} \overline{\langle \hat{V}_{R1} \hat{V}_{R2} \rangle}_R dt_1$.

According to the relation $e^A B e^{-A} = e^\nu B$ with $[A,B] = \nu B$, we have

$\hat{V}_R = \sum_q \gamma_q (e^{-i\Omega t}|0\rangle\langle 1| + e^{i\Omega t}|1\rangle\langle 0|)(e^{i\omega_q t} A_q^+ + e^{-i\omega_q t} A_q)$ and

$\hat{V}_A = -g(t) e^{i(\omega-\Omega)t}|0\rangle\langle 1| - g(t) e^{-i(\omega-\Omega)t}|1\rangle\langle 0|$.

Based on these equations we obtain

$$\begin{aligned}
\overline{\langle \hat{V}_{R1} \hat{V}_{R2} \rangle}_R &= \left\langle \hat{V}_R(t_1) \int_{-\infty}^{t_1} \hat{V}_R(t_2) dt_2 \right\rangle_R \\
&= \sum_{q,q'} \gamma_q \gamma_{q'} \int_{-\infty}^{t_1} dt_2 \langle (e^{-i\Omega t_1}|0\rangle\langle 1| + e^{i\Omega t_1}|1\rangle\langle 0|)(e^{i\omega_q t_1} A_q^+ + e^{-i\omega_q t_1} A_q). \\
&\quad (e^{-i\Omega t_2}|0\rangle\langle 1| + e^{i\Omega t_2}|1\rangle\langle 0|)(e^{i\omega_q t_2} A_{q'}^+ + e^{-i\omega_q t_2} A_{q'}) \rangle_R
\end{aligned} \tag{B-1}$$

According to Wick's theorem and energy conservation, equation (B-1) becomes

$$\begin{aligned}
\overline{\langle \hat{V}_{R1} \hat{V}_{R2} \rangle}_R &= \sum_q \gamma_q^2 \int_{-\infty}^{t_1} dt_2 [e^{i(\Omega-\omega_q-i\varepsilon)(t_2-t_1)} n_q |0\rangle\langle 0| \\
&\quad + e^{-i(\Omega-\omega_q+i\varepsilon)(t_2-t_1)} (1+n_q)|1\rangle\langle 1|], \\
&= i(\Gamma_1 |1\rangle\langle 1| - \Gamma_0 |0\rangle\langle 0|)
\end{aligned} \tag{B-2}$$

where $\Gamma_1 = \Gamma^*[1+N_B(\Omega)]$, $\Gamma_0 = \Gamma N_B(\Omega)$ and $\Gamma = \sigma + i\frac{\gamma}{2} = \sum_q \gamma_q^2 \frac{1}{\Omega - \omega_q - i\varepsilon}$. $N_B(\Omega)$ is the Bose-Einstein distribution function. Since the real part of $\Gamma_i$ just renormalizes the bare energy of state $|i\rangle$, equation (B-2) can be written as

$$\overline{\langle \hat{V}_{R1} \hat{V}_{R2} \rangle}_R = \gamma_0 |0\rangle\langle 0| + \gamma_1 |1\rangle\langle 1|, \tag{B-3}$$

with $\gamma_1 = \frac{\gamma}{2}[1+N_B(\Omega)]$, $\gamma_0 = \frac{\gamma}{2} N_B(\Omega)$. Then, we have $\hat{U}_I(t,t_\bullet)^+ = e^{-\sum_{j=0}^{1} \gamma_j |j\rangle\langle j|(t-t_\bullet)}$, and, thus,

$$\hat{U}(t_\bullet, t) = e^{iH_0(t-t_\bullet) - \sum_{j=0}^{1} \gamma_j |j\rangle\langle j|(t-t_\bullet)}. \tag{B-4}$$

Now, we can estimate $V_d$ and obtain

$$V_d = -ie^{-iH_0 t} \int_{-\infty}^{t} dt_1 \langle \hat{V}_R(t) \hat{V}_R(t_1) \rangle_R e^{iH_0 t} e^{-(t-t_1)\sum_{j=0}^{1} \gamma_j |j\rangle\langle j|}$$

$$+ (-i)^3 e^{-iH_0 t} \int_{-\infty}^{t} dt_1 \int_{-\infty}^{t_1} dt_2 \int_{-\infty}^{t_2} dt_3 \langle \hat{V}_R(t) \hat{V}_A(t_1) \hat{V}_A(t_2) \hat{V}_R(t_3) \rangle_R e^{iH_0 t} e^{-(t-t_3)\sum_{j=0}^{1} \gamma_j |j\rangle\langle j|}. \tag{B-5}$$

$$+ \cdots$$

Calculating the last one integral in each term, yields

$$V_d = [\Gamma_1' |1\rangle\langle 1| - \Gamma_0' |0\rangle\langle 0|]$$
$$- \left\{ \Gamma_1' \int_{-\infty}^{t} dt_1 \int_{-\infty}^{t_1} dt_2 g(t_1) e^{i(\omega-\Omega)t_1} g(t_2) e^{-i(\omega-\Omega)t_2} e^{\gamma_1(t_2-t)} |1\rangle\langle 1| \right.$$
$$\left. - \Gamma_0' \int_{-\infty}^{t} dt_1 \int_{-\infty}^{t_1} dt_2 g(t_1) e^{-i(\omega-\Omega)t_1} g(t_2) e^{i(\omega-\Omega)t_2} e^{\gamma_0(t_2-t)} |0\rangle\langle 0| \right\}, \tag{B-6}$$
$$+ - \cdots$$

where $\Gamma_0' = \sum_q \gamma_q^2 \frac{1}{\Omega - \omega_q - i\gamma_0} N_B(\Omega) \cong \Gamma_0$ and $\Gamma_1' = \sum_q \gamma_q^2 \frac{1}{\Omega - \omega_q + i\gamma_1}[1+N_B(\Omega)] \cong \Gamma_1$. For a square-shape pulse $g(t) = \lambda[\theta(t-t_0) - \theta(t-t_0-\tau_{on})]$ with pulse height $\lambda$, pulse width $\tau_{on}$, and starting time $t_0$, and integrating equation (B-6) term-by-term in the case of $\omega = \Omega$ and $t_0 \leq t \leq t_0 + \tau_{on}$, we find that equation (B-6) can be written as

$$V_d = \Gamma_1|1\rangle\langle 1|\left\{1 - \frac{\lambda\theta(\tau)}{\gamma_1}\int_0^{\gamma_1\tau} dy e^{-y}[\frac{1}{1!}\frac{y\lambda}{\gamma_1} - \frac{1}{3!}(\frac{y\lambda}{\gamma_1})^3 + \cdots]\right\}$$

$$-\Gamma_0|0\rangle\langle 0|\left\{1 - \frac{\lambda\theta(\tau)}{\gamma_0}\int_0^{\gamma_0\tau} dy e^{-y}[\frac{1}{1!}\frac{y\lambda}{\gamma_0} - \frac{1}{3!}(\frac{y\lambda}{\gamma_0})^3 + \cdots]\right\}$$

$$= -i\gamma_1|1\rangle\langle 1|\left\{1 + \frac{\lambda e^{-\gamma_1\tau}[\lambda\cos(\lambda\tau) + \gamma_1\sin(\lambda\tau)] - \lambda^2}{\gamma_1^2 + \lambda^2}\theta(\tau)\right\}$$

$$-i\gamma_0|0\rangle\langle 0|\left\{1 + \frac{\lambda e^{-\gamma_0\tau}[\lambda\cos(\lambda\tau) + \gamma_0\sin(\lambda\tau)] - \lambda^2}{\gamma_0^2 + \lambda^2}\theta(\tau)\right\}$$

(B-7)

where $\tau = t - t_0$. Similarly, for $t > t_0 + \tau_{on}$, we obtain

$$V_d = -i\gamma_1|1\rangle\langle 1|\left\{1 - e^{-\gamma_1\tau}\frac{\lambda[\lambda e^{\gamma_1\tau_{on}} - \lambda\cos(\lambda\tau_{on}) - \gamma_1\sin(\lambda\tau_{on})]}{\gamma_1^2 + \lambda^2}\theta(\tau)\right\}$$

$$-i\gamma_0|0\rangle\langle 0|\left\{1 - e^{-\gamma_0\tau}\frac{\lambda[\lambda e^{\gamma_0\tau_{on}} - \lambda\cos(\lambda\tau_{on}) - \gamma_0\sin(\lambda\tau_{on})]}{\gamma_0^2 + \lambda^2}\theta(\tau)\right\}$$

(B-8)

For the same derivation procedure, for $\omega = \Omega$ and $t_0 \leq t \leq t_0 + \tau_{on}$, $V_{od}$ has the form

$$V_{od} = i\Gamma_1|0\rangle\langle 1|e^{i\Omega t}\frac{\lambda\theta(\tau)}{\gamma_1}\int_0^{\gamma_1\tau} dy e^{-y}\cos(\frac{\lambda y}{\gamma_1}) - i\Gamma_0|1\rangle\langle 0|e^{-i\Omega t}\frac{\lambda\theta(\tau)}{\gamma_0}\int_0^{\gamma_0\tau} dy e^{-y}\cos(\frac{\lambda y}{\gamma_0})$$

$$= i\Gamma_1\lambda|0\rangle\langle 1|e^{i\Omega t}\frac{\gamma_1 + e^{-\gamma_1\tau}[\lambda\sin(\lambda\tau) - \gamma_1\cos(\lambda\tau)]}{\gamma_1^2 + \lambda^2}\theta(\tau)$$

$$-i\Gamma_0\lambda|1\rangle\langle 0|e^{-i\Omega t}\frac{\gamma_0 + e^{-\gamma_0\tau}[\lambda\sin(\lambda\tau) - \gamma_0\cos(\lambda\tau)]}{\gamma_0^2 + \lambda^2}\theta(\tau)$$

(B-9)

For $t > t_0 + \tau_{on}$, we obtain

$$V_{od} = \gamma_1\lambda|0\rangle\langle 1|e^{i\Omega t}e^{-\gamma_1\tau}\frac{\gamma_1 e^{\gamma_1\tau_{on}} - \gamma_1\cos(\lambda\tau_{on}) + \lambda\sin(\lambda\tau_{on})}{\gamma_1^2 + \lambda^2}\theta(\tau)$$

$$+\gamma_0\lambda|1\rangle\langle 0|e^{-i\Omega t}e^{-\gamma_0\tau}\frac{\gamma_0 e^{\gamma_0\tau_{on}} - \gamma_0\cos(\lambda\tau_{on}) + \lambda\sin(\lambda\tau_{on})}{\gamma_0^2 + \lambda^2}\theta(\tau)$$

(B-10)

The results of $V_d$ and $V_{od}$ for N identical pulses are shown in equations (11) and (12) of the text.

For large detuning ($\delta\omega \gg \lambda$) and $t_0 \leq t \leq t_0 + \tau_{on}$, equation (B-6) has the approximate form

$$V_d \cong -i \sum_{j=0}^{1} \gamma_j \left\{ 1 + \frac{(-1)^j i \lambda^2}{\delta \omega} \right.$$

$$\left. \times \int_0^{\gamma_j \tau} dy\, e^{-y} \left[ 1 + \frac{(-1)^j i \lambda^2 y}{\gamma_j \delta \omega} + \frac{1}{2!}\left(\frac{(-1)^j i \lambda^2 y}{\gamma_j \delta \omega}\right)^2 + \cdots \right] \theta(\tau) \right\} |j\rangle\langle j| \quad . \tag{B-11}$$

$$= -i\gamma_j \sum_{j=0}^{1} \left[ 1 + \frac{\lambda^2}{\lambda^2 + (-1)^j i \gamma_j \delta \omega} e^{-\gamma_j \tau}\left(e^{(-1)^j i \lambda^2 \tau/\delta\omega} - e^{\gamma_j \tau}\right) \theta(\tau) \right] |j\rangle\langle j|$$

For $t > t_0 + \tau_{on}$, we have

$$V_d \cong -i \sum_{j=0}^{1} \gamma_j \left\{ 1 + \frac{(-1)^j i \lambda^2}{\delta \omega} e^{-\gamma_j (\tau - \tau_{on})} \right.$$

$$\left. \times \int_0^{\gamma_j \tau_{on}} dy\, e^{-y} \left[ 1 + \frac{(-1)^j i \lambda^2 y}{\gamma_j \delta \omega} + \frac{1}{2!}\left(\frac{(-1)^j i \lambda^2 y}{\gamma_j \delta \omega}\right)^2 + \cdots \right] \theta(\tau) \right\} |j\rangle\langle j| \quad . \tag{B-12}$$

$$= -i \sum_{j=0}^{1} \gamma_j \left[ 1 + \frac{\lambda^2}{\lambda^2 + (-1)^j i \gamma_j \delta \omega} e^{-\gamma_j \tau}\left(e^{(-1)^j i \lambda^2 \tau_{on}/\delta\omega} - e^{\gamma_j \tau_{on}}\right) \theta(\tau) \right] |j\rangle\langle j|$$

The result of $V_d$ for N identical pulses is shown in equation (13) of the text.